\documentclass[11pt]{aa}
\usepackage[english]{babel}
\usepackage{graphicx}
\usepackage{longtable}

\onecolumn
\begin{document}

\title{Extended atmosphere of the yellow hypergiant V509\,Cas \\  in 1996--2018}

\author{V.G.~Klochkova, E.L.~Chentsov, V.E.~Panchuk}
\institute{Special Astrophysical Observatory RAS, Nizhnij Arkhyz,  369167 Russia}

\date{\today} 

\abstract{Based on the data of spectral monitoring of the yellow hypergiant V509\,Cas performed in
1996--2018 at the 6-m telescope with the spectral resolution of R$\ge$60\,000, we studied in detail its
kinematic state at various levels of extended atmosphere. No signs of presence of a companion were found.
An agreement of radial velocities measured on the permitted and forbidden emissions of metal ions, as well
as their strict temporal stability led us to the choice of the systemic velocity of the star Vsys=$-63$\,km/s.
The position of forbidden [NII] emissions forming in the circumstellar medium is strictly stable and is
systematically shifted by $-6$\,km/s relative to the metal ion emissions. A conclusion on the variation of
the [NII] emission halfwidths and intensities (the lines have become narrower and more intense) is made
after the observations in 1996 and these parameters did not vary over the next 22 years of observations. 
The velocities measured from the shortwave FeII(42) absorption components are located in a narrow interval
of Vr=$-$(84$\div$87)\,km/s, which indicates the stability of expansion of the upper layers of the atmosphere.
The overall atmosphere of the hypergiant is stable, excluding the layers close to the photosphere. The
velocity variability in range of Vr=$-$(52$\div$71)\,km/s, identified by the positions of strong metal ion
absorption cores, may be a manifestation of pulsations in deep atmospheric layers, where this type of lines
are formed.
}

\titlerunning{\it Extended atmosphere of V509\,Cas hypergiant in 1996--2018}  
\authorrunning{\it Klochkova et al.}

\maketitle

\section{Introduction}

A yellow hypergiant V509\,Cas  is a prominent representative of a group of rare stars near the luminosity limit. 
The general data about these objects with the initial masses of 20$\div$40$\mathcal{M}_{\odot}$ 
are represented in~[1, 2]. In addition to their extreme luminosity, yellow hypergiants differ from ordinary
supergiants by a very high mass-loss rate in the stellar wind and the presence of gas and dust circumstellar envelope. 
Hypergiant instability is manifested as a weak brightness variability (with an amplitude of  $\approx 0.2\div0.4^m$), 
which is usually referred to as the pulsation type. Along with the manifestations of instability, yellow hypergiants 
also experience sporadic pulsations, the so-called ``shell episodes'', during which the star is particularly intensively 
losing its mass and for several hundreds of days gets enveloped by the ejected cold matter forming a pseudophotosphere.

Yellow hypergiants occupy a limited area on the Hertzsprung--Russell diagram~[3]. However, observable properties of 
well-studied representatives of this family ($\rho$\,Cas, V509\,Cas, V1302\,Aql) significantly differ. This has to 
do primarily with the differences in the optical spectra and the features of the structure and kinematics of the envelopes. 
For example, in the spectrum of the yellow hypergiant $\rho$\,Cas, which is considered as a prototype of a group of yellow
hypergiants, a significant variability of the emission-absorption profile H$\alpha$, as well as the splitting of the
BaII, SrII, TiII and other strongest absorptions with a low excitation potential of the lower level was
discovered more than half a century ago by Bidelman and McKellar~[4]. Later, these features of the $\rho$\,Cas
spectrum were studied in detail by spectral monitoring~[5--9]. On the HR diagram $\rho$\,Cas is located at the boundary of 
the Yellow Void~[2], separating the hypergiants and LBVs in the quiet phase. On the boundary of the Yellow Void, 
the amplitude of pulsations of yellow hypergiants apparently sharply increases, what leads to an increased instability of
the atmosphere and a blowing off of the envelope~[2]. In 2013, the $\rho$\,Cas system underwent a new mass ejection, 
at which the star’s brightness dropped by 0.5$^m$. This ejection occurred only 12 years after the previous one in 2000--2001. 
This way, there is an increase in ejections in $\rho$\,Cas, which, according to Aret et al.~[10] may suggest an approach 
of the star to the boundary of the Yellow Void. Surprisingly, with such a vigorous mass loss, an optical spectrum so
rich in features and variability, this hypergiant does not have a circumstellar dust envelope.

A close relative of $\rho$\,Cas, a yellow hypergiant V1302\,Aql, is on the contrary associated with one of the brightest 
sources of IR-radiation IRC+10420. In the system of this object a source of maser radiation in OH was also detected~[11]. 
The optical spectrum of V1302\,Aql is dominated by the emission and absorption lines of metal ions FeII, TiII, ScII, CrII,
emission lines FeI and absorption lines NI, OI and SiII, as well as the forbidden lines [FeII], [CaII]
and [OI]~[12--15]. Metal ion line profiles are very diverse: from almost symmetrical emissions to the reverse P\,Cyg-type  
profiles and absorption profiles with two emission components. A long-term spectral monitoring of V1302\,Aql, 
performed during the past decades~[12, 14, 15], led to the conclusion that the object had passed on the HR-diagram the path from
the red supergiant to the cold border of the Yellow Void.

The difficulty of studying the spectra of yellow hypergiants is well-illustrated by the history of the study
of the cool supergiant HD\,179821, identified with an IR source IRAS 19114+0002. A combination of the
observed parameters of this star for a long time did not allow to unambiguously determine its evolutionary
status (see the paper~[16] and references therein). However, over the past years, strong enough evidence
for the belonging of the star to yellow hypergiants was obtained~[17, 18].
The supergiant V509\,Cas, like V1302\,Aql, described a complex path, a ``zigzag'' on the HR-diagram,
as expressed in~[3]. In the optical wavelength range the spectrum of this hypergiant varied at different
points of observation in the range of spectral classes G--K, while preserving the Ia luminosity class~[19].
The optical spectrum of V509\,Cas is rich in emission features that distort the absorption profiles not only
of hydrogen, but also of metallic lines, which in one way or another is inherent in the spectra of other studied
yellow hypergiants. At the same time, the spectrum of V509\,Cas revealed a unique feature--emissions
of highly excited forbidden lines of [NII]\,6548 and  6584\,\AA{}  already found in the spectrum of this star
in 1961~[20], when this star was considered as an MK-classification standard. Their presence in the
spectrum of such a cool single star is difficult to explain. This fact served as an incentive for us to
conduct a long-term monitoring of the star with high spectral resolution.

In this paper, we present the results obtained based on the monitoring of the spectrum of the star
over several sets in 1996--2018. Section~2 briefly  describes the methods of observations and data
analysis. In Section~3 we present the results, compare them with the ones previously published. Section~4
gives the conclusions.

\section{Observations, reduction and analysis of spectra}

We have carried out all observations of V509\,Cas with the echelle spectrographs in the Nasmyth focus of the 
6-m BTA telescope of the Special Astrophysical Observatory of the RAS. The dates of
observations and registered spectral ranges are given in Table~1. In 1996 we used the Lynx echelle
spectrograph~[21], which in combination with the CCD chip 1K$\times$1K provided the spectral resolution of
R$\approx$25\,000. All subsequent spectral data were obtained with the NES echelle spectrograph~[22, 23],
equipped with a large format CCD sized 2048$\times$4608 pixels. To reduce light loss without a 
deterioration of spectral resolution, the NES spectrograph is equipped with an image slicer (which provides three
slices of a stellar image). Every spectral order on the image is repeated three times with an offset along the 
dispersion of an echelle grating~[23]. A transition to the large format CCD substantially extended the boundaries
of the simultaneously recorded wavelength interval: for example, $\Delta\lambda$=5400$\div$8479\,\AA{} in the spectrum
of 1.10.2014 or $\Delta\lambda$=4697$\div$7782\,\AA{} in the spectra of 2017. Spectral resolution is 
$\lambda/\Delta\lambda\ge$60\,000, the signal-to-noise ratio is S/N$>$100 and differs little
from one spectrum to another.

\begin{table*}[t!]
\medskip
\caption{Heliocentric radial velocity Vr measurements, rounded to the nearest km/s, in the spectra of V509\,Cas from
          sets of various types of lines}
\begin{tabular}{ c|  c|  c|  c  c  c|  c| c| c| c| c| c| c| c  }
\hline
Date  &$\Delta\lambda$, &\multicolumn{12}{c}{\small  Vr, km/s} \\
\cline{3-14}
    & nm  & [FeII]  &\multicolumn{3}{c|}{\small lines of ions}& \small FeII(42) &\small H$\alpha$& [NII]&\multicolumn{4}{c|}{\small NaI} & DIBs \\
\cline{10-13}
                      &  &  emis & emis & wing   &  core &  abs   & abs & &\small  CS & st &  IS& IS &   \\   
\hline   
1 & 2 & 3 & 4 & 5 & 6 & 7 & 8 & 9 & 10 & 11 & 12 & 13 & 14 \\
\hline
 02.05.1996 & 527-685 & $-62$&$-62$ &$-62$ & $-71$&           &$-104,-11$&$-69$ &$-89$&$-62$&$-47$&$-15$&$-14$\\  
 03.07.1996 & 515-800 & $-63$&$-62$ &$-63$ & $-66$&           &$-103,-20$&$-68$ &$-88$&$-62$&$-48$&$-14$&$-15$\\  
 01.10.2014 & 540-848 & $-63$&$-63$ &$-60$ & $-53$&           &$ -96,-18$&$-69$ &$-90$&$-62$&$-50$&$-14$&$-13$\\  
 04.09.2015 & 395-666 & $-62$&$-63$ &$-62$ & $-60$&$-87, -55$ &$ -96,-22$&$-69$ &$-90$&$-62$&$-51$&$-14$&$-15$\\  
 26.10.2015 & 398-676 & $-63$&$-63$ &$-63$ & $-62$&$-87, -53$ &$ -95,-22$&$-68$ &$-90$&$-62$&$-50$&$-14$&$-14$\\  
 12.02.2017 & 470-778 & $-62$&$-63$ &$-62$ & $-59$&$-84, -52$ &$ -93,-27$&$-69$ &$-89$&$-62$&$-50$&$-13$&$-14$\\  
 13.06.2017 & 470-778 & $-63$&$-63$ &$-63$ & $-62$&$-86, -54$ &$ -92,-26$&$-69$ &$-89$&$-62$&$-50$&$-14$&$-14$\\  
 03.08.2017 & 470-778 & $-63$&$-64$ &$-63$ & $-62$&$-85, -54$ &$ -92,-21$&$-69$ &$-89$&$-62$&$-52$&$-14$&$-14$\\  
 06.04.2018 & 470-778 & $-63$&$-62$ &$-61$ & $-59$&$-84, -55$ &$ -93,-20$&$-68$ &$-89$&$-61$&$-52$&$-14$&$-14$\\  
\hline
\end{tabular}   
\end{table*}

Extraction of one-dimensional data from the twodimensional echelle spectra was performed using an
ECHELLE context from the MIDAS software package modified taking into account the features of the
echelle frames of the spectrographs used (see the features in~[24]). The removal of traces of cosmic
particles was carried out by the median averaging of two spectra, obtained sequentially one after the other.
The wavelength calibration was performed based on the spectra of a Th--Ar hollow-cathode lamp. We
identified the features in the spectrum of V509\,Cas using the previously published spectral atlas [25].
All further reduction, including the photometric and positional measurements was done using the
latest version of the DECH20t code~[26]. Note that this spectrum reduction program we traditionally use
allows to measure radial velocities for individual features of line profiles, and that the paper only uses
the heliocentric velocities Vr. Systematic errors in measuring Vr, estimated by sharp interstellar components 
of NaI do not exceed 0.25\,km/s (from one line), a random error for shallow absorptions of about 0.7\,km/s  -- 
the average value per line. Thereby for our averaged values in Table~1 random errors are about 0.3\,km/s.

\section{Discussion of results} 

\subsection{Main features of the optical spectrum of V509\,Cas}

V509\,Cas is a very attractive object for a detailed high-resolution spectroscopy: this bright star has coordinates, 
permitting year-round observations in the northern hemisphere, and its spectrum is abound in a variety of complex features, 
which can be clearly seen in Figs.\,1$\div$3. Naturally, such an interesting object has a long history of spectral 
observations, and the main features of its spectrum were found in the earliest publications~[20]. Since 1974, its spectral 
class has become earlier: in 1977 it was classified as F8 (the same as in $\rho$\,Cas~[19]). According to our observations,
from 1996 to 2018 the spectrum of V509\,Cas varied little, and the spectral class remained even earlier,
close to F2. We made this estimate visually from a comparison of the spectrum of V509\,Cas with the
data from the spectral atlas of F- and G-stars~[25] and quantitatively from the FeI and FeII line intensity ratio. 
Figure~4 shows the evolution of the stellar spectrum with time on its small fragment with the FeI and FeII lines. 
Two upper fragments are borrowed from the above work~[19]. This figure indicates the absence of significant changes 
of spectral class in V509\,Cas during our observations over 1996--2018, which is consistent with the constancy of 
the effective temperature of the star during these years (see Fig.\,2 in~[3]).

As noted in the Introduction, in terms of its effective temperature and its evolutionary status V509\,Cas
is close to a related object V1302\,Aql. The optical spectrum of the hypergiant V1302\,Aql is dominated
by the emission and absorption lines of FeII, TiII, ScII, CrII metal ions, FeI emission lines and NI,
OI and SiII absorptions~[14, 27]. Some emission features are identified with the forbidden lines of
[FeII], [CaII] and [OI]. The metal ion line profiles are very diverse: from the almost symmetrical emissions 
to the reverse P\,Cyg-like profiles and absorption profiles with two emission components. The high excitation 
lines (NI, OI and SiII) have almost symmetrical absorption profiles, shifted on the average by
5\,km/s red-wave relative to the positions of minima of their profiles. The spectra of V509\,Cas reveal
roughly the same set of metal lines, but the FeI lines are more often purely absorption, or absorption with
emissions in the wings.

The hydrogen H$\alpha$ and H$\beta$ lines in the spectra of V1302\,Aql have a characteristic two-peak profile,
similar to the H$\alpha$ profile in the spectrum of V509\,Cas. But compared to the spectrum of V509 Cas, the ratio 
of emission peaks in the spectrum of V1302\,Aql is the opposite: the short-wave peak is much more
intense than the long-wave peak. Only on one of the dates (11/24/2007) of many years of monitoring of
V1302\,Aql the 6-m telescope recorded an unusual view of the H$\alpha$ profile, in which the longwave peak 
substantially exceeds the shortwave one. In addition, unlike V509\,Cas, in the spectrum of V1302\,Aql a
broad photospheric absorption is completely flooded with the emission. 
For a more detailed comparison of the spectra of two hypergiants it is convenient to use
the atlas of the V1302\,Aql spectrum~[28].

\begin{figure*}[t!]
\vbox{
\includegraphics[angle=0,width=0.5\textwidth, bb=30 70 770 525,clip]{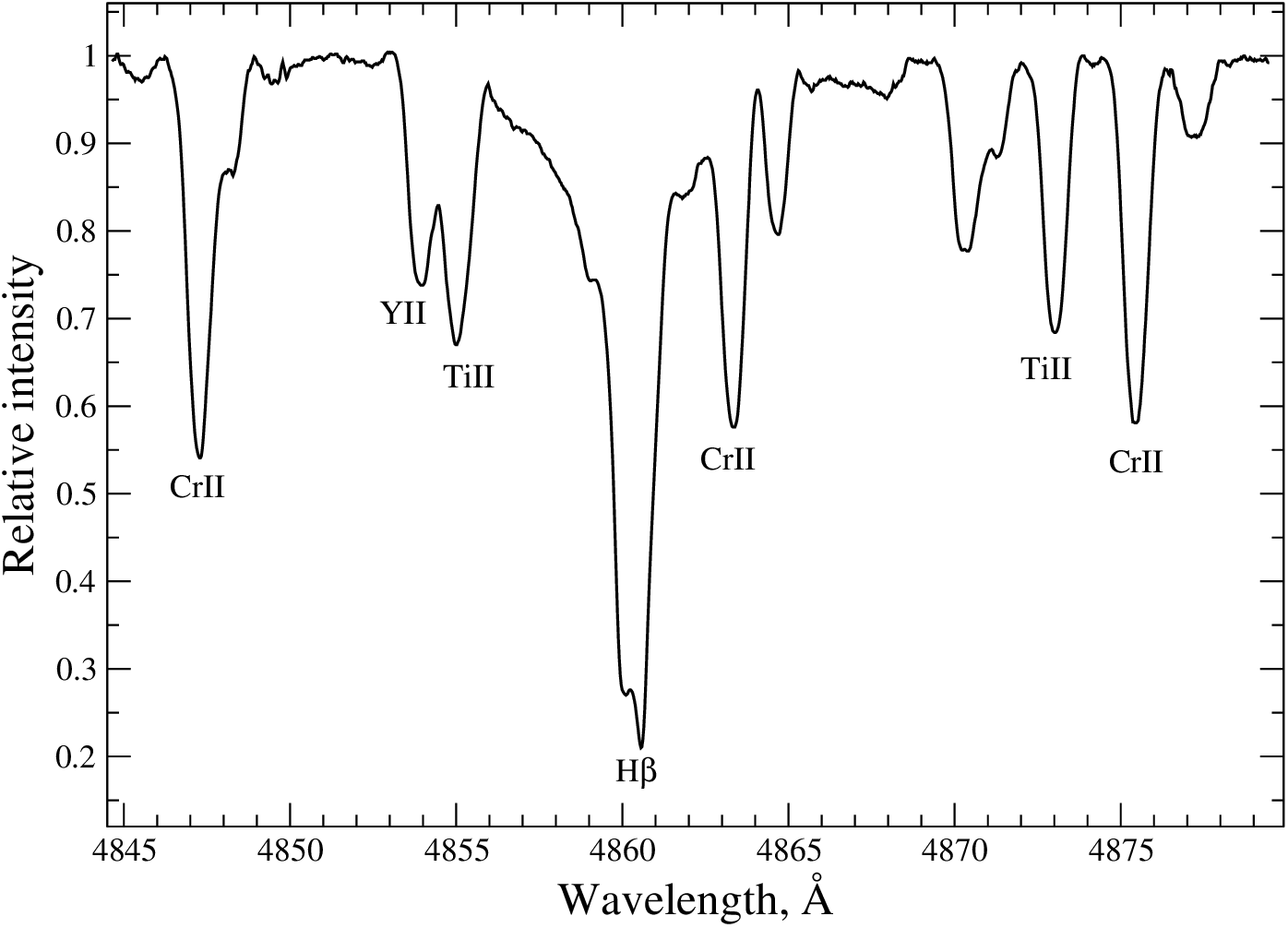} 
\includegraphics[angle=0,width=0.5\textwidth, bb=30 70 770 525,clip]{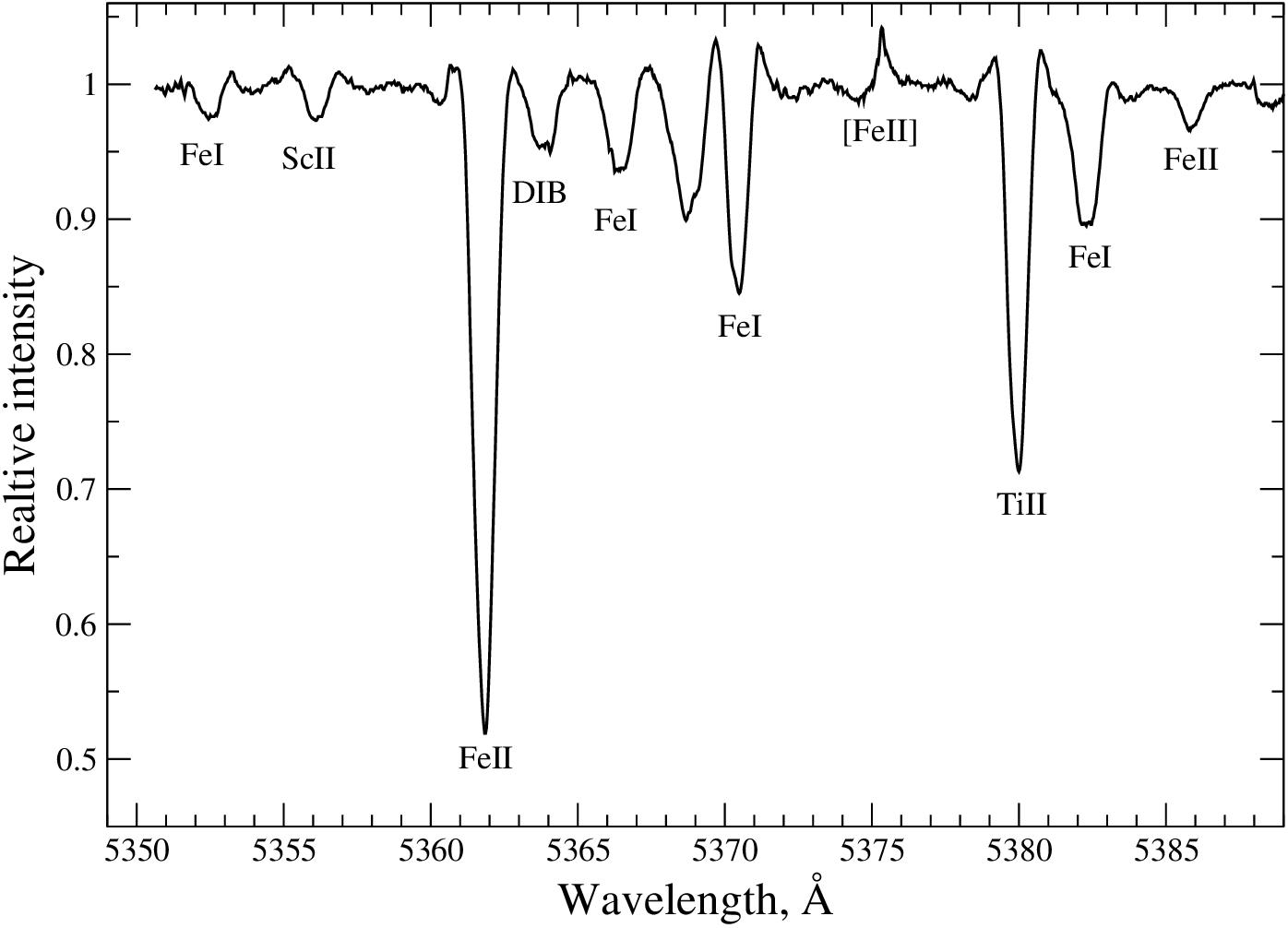}
\includegraphics[angle=0,width=0.5\textwidth, bb=30 40 770 525,clip]{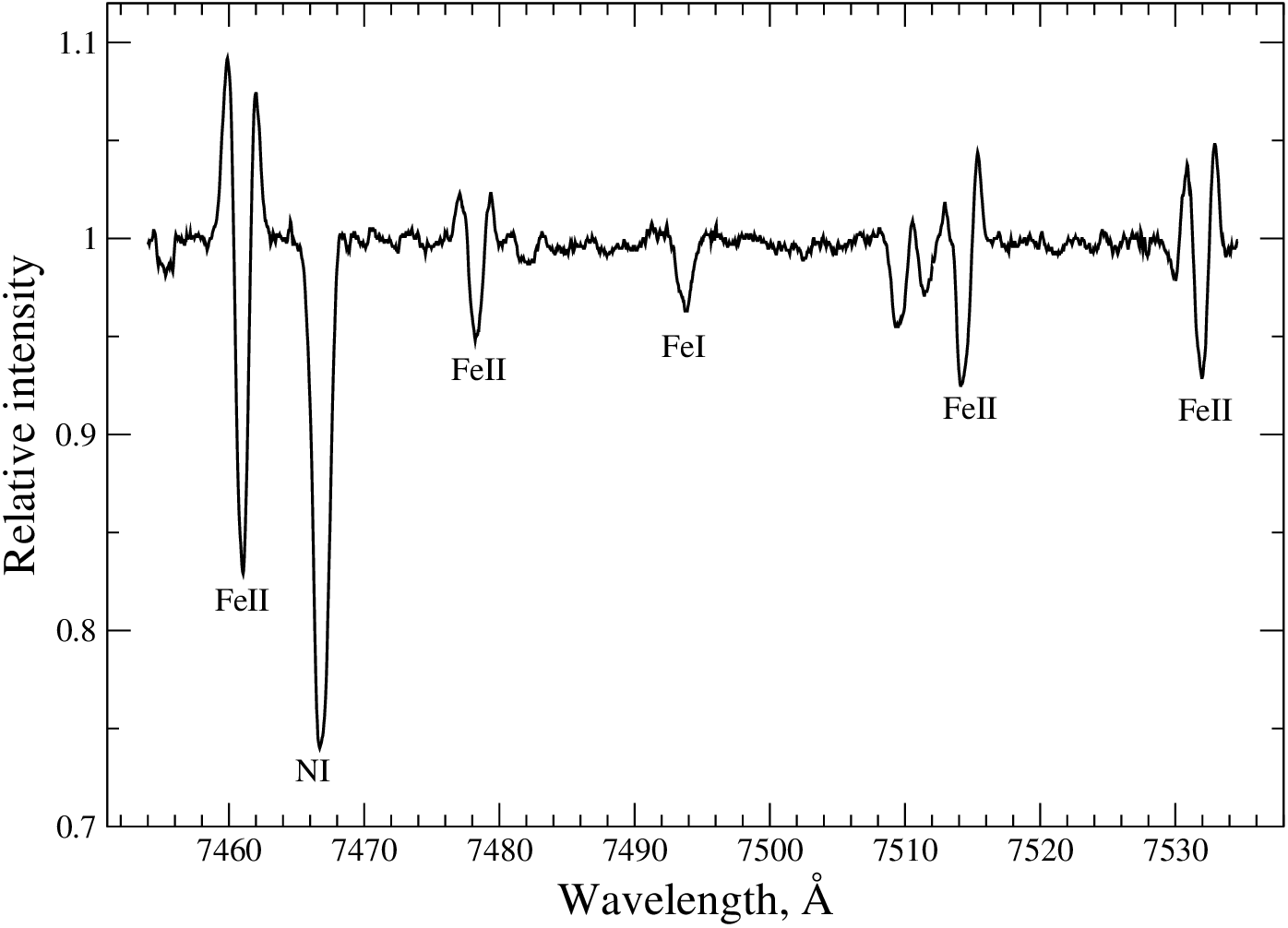} 
}
\caption{Fragments of the spectrum of V509 Cas: (a) the interval of 4845$\div$4880\,\AA{}, 
      containing the H$\beta$ line; (b) the interval of 5350$\div$5390\,\AA{} with the forbidden 
      line [FeII]\,5376\,\AA{}; (c) the interval of 7455$\div$7535\,\AA{} with the NI\,7468\,\AA{} line. 
      The maiin features are identified.}
\end{figure*}

\begin{figure*}[t!]
\includegraphics[angle=0,width=0.8\textwidth, bb=70 10 800 610,clip]{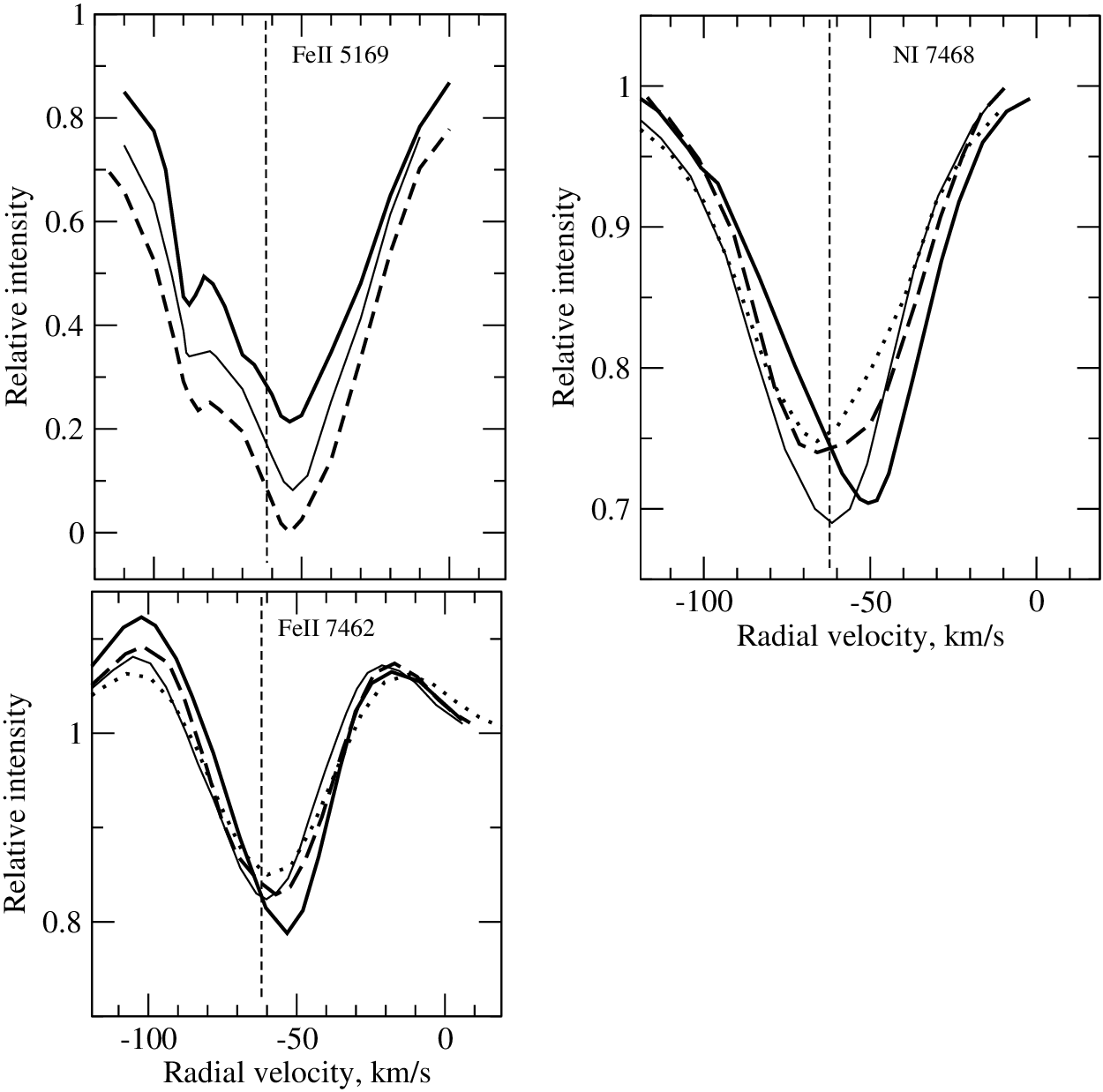}   
\caption{The variations of line profiles in the spectra of V509\,Cas obtained over different years: 
   the dotted line  -- in 1996, the solid thick line --- in 2014, the solid thin line -- in 2017, the dashed 
   line -- in 2018. The FeII\,5169\,\AA{} line profiles are shifted along the ordinate axis with respect to 
   the previous profile by 0.1. The vertical dashed line indicates the accepted value of systemic velocity   
   Vsys\,=\,$-63$\,km/s.} 
\end{figure*}

All our spectra of V509\,Cas contain the H$\alpha$ line, some of its profiles are presented in Fig.\,3, 
from which stems an insignificant profile variability over the whole period of 1996--2018. The intensity of
the shortwave emission peak is constantly below the longwave one. Mainly the shortwave absorption
wing is exposed to the wind-induced variability. There are no noticeably significant changes either
in the shapes of metal line profiles or in the positions of their main components. This is shown in
Fig.\,2, which demonstrates the absorption profiles of NI\,7468 and FeII\,5169\,AA{} (fragments with the
profile of FeII\,5169\,\AA{} are shifted vertically relative to the top one) the absorption-emission profile of the
FeII\,7462\,\AA{} line, as well as Table~1.

V509\,Cas is usually considered to be a spectral twin of the hypergiant  $\rho$\,Cas~[29], and therefore many
authors compare their spectra~[5, 29$\div$31]. However, a detailed study of the spectra of both hypergiants
with close fundamental parameters (mass, luminosity, evolutional stage) revealed significant differences
both in the spectra and kinematic conditions of their atmospheres, indicating differences in physical processes 
causing an instability of the atmospheres and envelopes. Spectral differences are manifested primarily in the 
differences of the H$\alpha$ profile and its variability. 

The H$\alpha$ profile in the spectrum of $\rho$\,Cas significantly varies mainly due to the envelope ejections.
The H$\alpha$ profile variations are especially demonstrative after the 2013 ejection~[9]. The shifts of the H$\alpha$ 
profile indicate a change in structure and increased instability of the upper layers of the extended atmosphere
of the star. A significant shift of the profile to the long-wavelength region in the spectra of August--September 
2017 indicates the infall  of the layers of matter where the line is formed. 

As follows from Fig.\,3, the behavior  of H$\alpha$ profile in the spectrum of V509\,Cas is according to our observations 
more calm. The H$\alpha$ absorption components in this figure and FeII\,(42) lines in Fig.\,2 are located on both sides of 
the Vsys line, which indicates the presence in the atmosphere and in the envelope of V509\,Cas of both the outflowing and 
accreting matter. 
The profile variations are observed in the shift and variable width of the absorption component (Fig.\,3) caused by 
unstable conditions in the envelope, expanding at a velocity of 33$\div$40\,km/s. Note that the authors of~[30], comparing 
the spectra of V509\,Cas and  $\rho$\,Cas in the near UV, have also emphasized that these two stars are not spectral
twins.

\begin{figure*}[t!]
\includegraphics[angle=0,width=0.7\textwidth, bb=40 55 610 530,clip]{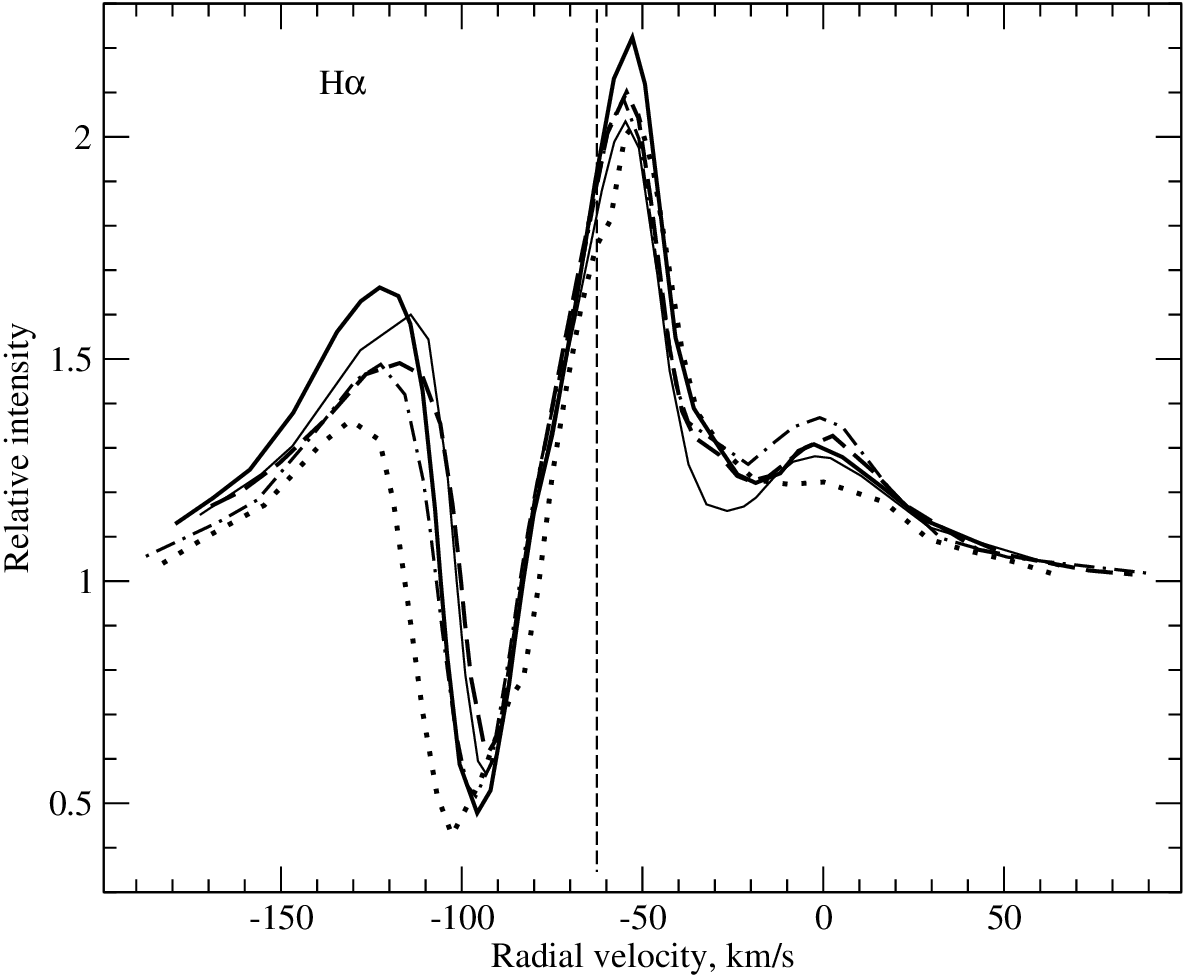}   
\caption{The variations of the H$\alpha$ profile in the spectra of V509\,Cas obtained on various dates: 05/02/1996 -- the dotted
     line, 10/01/2014 -- the thick solid line, 10/26/2015 -- the dash-dotted line, 2017 -- the thin solid line, 04/06/2018 -- the
     dashed line. Vertical dashed line indicates the accepted value of the systemic velocity Vsys\,=\,$-63$\,km/s.} 
\end{figure*}

\begin{figure*}[t!]
\includegraphics[angle=0,width=0.7\textwidth, bb=35 40 710 530,clip]{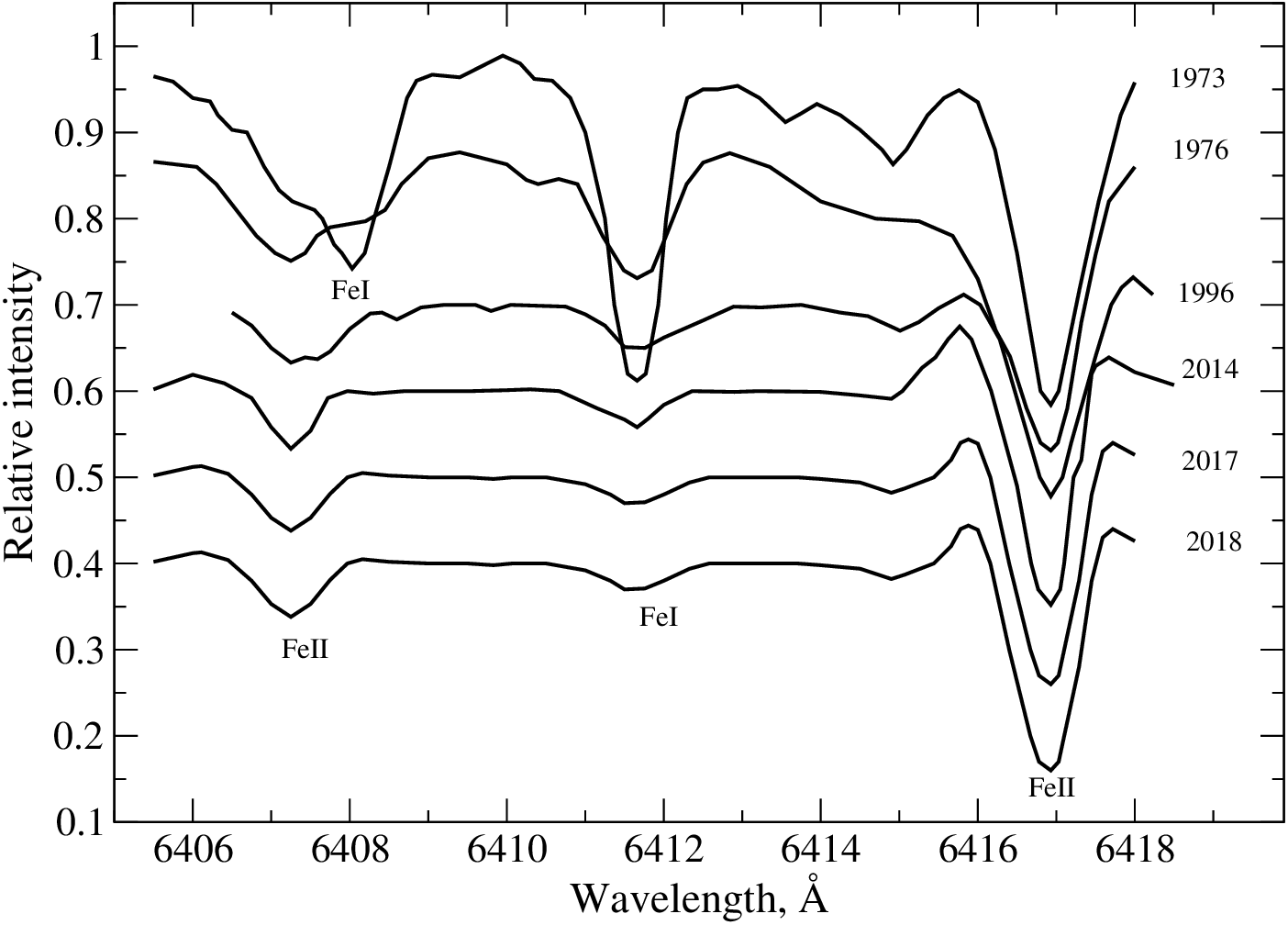}   
\caption{Variability of the spectrum of V509\,Cas in the wavelength range of 6405--6418\,\AA{}. 
    The two upper curves are adopted from~[19].} 
\end{figure*}

In addition, a temporarily variable splitting of the strongest absorptions and envelope emissions of the
iron group atoms is observed in the spectrum of  $\rho$\,Cas (see [9] and the references therein). Both of
these effects due to the presence of a powerful structured envelope in $\rho$\,Cas, are absent in the spectrum
of V509\,Cas. On the other hand, the spectrum of V509\,Cas possesses the features missing in the spectrum of $\rho$\,Cas: 
the highly excited forbidden lines [NII] and emission components in most of the permitted and forbidden metal ion 
lines (see article~[20] and Fig.\,1 and 4 in this publication). A number of diffuse interstellar
features are identified in the spectrum of V509\,Cas, while they are absent in the spectrum of $\rho$\,Cas, which
has close galactic coordinates. 

The spectrum of V509\,Cas has a similarity with the spectrum of a hotter high-luminosity star 3\,Pup (Sp\,=\,A4\,Iabe). 
In the spectrum of this massive supergiant, studied in detail in~[32], a bifurcated H$\alpha$
line was registered with a stronger red component (a P\,Cyg\,III-type profile according to Beals~[33]), the
emissions of forbidden [OI]\,1F\,6300, 6364\,\AA{} lines, the doublet [CaII]\,1F\,7291, 7324\,\AA{} and emissions in
some Fe\,II lines. The authors of~[32] stressed that in the spectrum of 3\,Pup only the MgII\,4481\,\AA{} line
can be considered photospheric, and in FeII lines a contribution of the envelope is obvious, which gives
the profiles a specific shape: the wings are raised by the emissions. The same type of features we also
observe in the spectrum of V509\,Cas, in which the wings of even the weak lines are distorted by the
emissions, which is clearly seen in Fig.\,2 for the FeII\,7462\,\AA{} line with a depth of not more than 0.2
from the level of the local continuum. The star 3\,Pup is the coldest member of the family of supergiants
with the B[e] phenomenon, discussed in detail in the paper of Aret et al.~[34].

\subsection{Forbidden lines of [NII]}

The presence of various emission combinations [OI], [OII], [NII], [SII], [CaII], etc. is not a sign
of any certain class of stars. Strong forbidden emissions whose region of formation is a dense rotating disk~[34] 
are ordinary features in the spectra of stars with the B[e] phenomenon. An example of this class of stars are the well known 
star  CI\,Cam   and above-mentioned  3 Pup, in the spectrum of which there are many forbidden emissions, but the [NII] 
emissions are absent. The presence of forbidden emissions is known in other types of hot, far evolved stars that have passed
stages with a substantial loss of matter: a hypergiant MWC\,314~[28], an LBV-star GR-290~[35], an X-ray transient~[36, 37], 
a protoplanetary nebulae that entered the envelope ionization phase [38--42]. The objects with forbidden emissions of 
light metals and their ions are often binary systems with a hot component~[43].

\begin{table}[t!]
\medskip
\caption{The results of measurements of the parameters of [NII] lines in the spectra of V509\,Cas. 
  Half-widths of $\delta$Vr profiles are rounded to whole km/s, the equivalent  widths W$_{\lambda}$ 
  are given in \AA{}. Uncertain values are marked by the colon.} 
\begin{tabular}{ r| c c| c  c| c  c} 
\hline
Date   &\multicolumn{2}{c|}{\small[NII]\,5755}& \multicolumn{2}{c|}{\small[NII]\,6548} & \multicolumn{2}{c}{\small [NII]\,6583} \\
  & $\delta$Vr & W$_{\lambda}$  &  $\delta$Vr & W$_{\lambda}$  & $\delta$Vr & W$_{\lambda}$ \\   
\hline
1961$^1$  &   &       &    &      & 50&       \\ 
1967$^2$  &   &       &    & 0.045& 30& 0.13  \\ 
2.05.1996 & 20& 0.025 & 30 &0.10  & 29& 0.33  \\ 
3.07.1996 & 19& 0.026 & 32 &0.17: & 28& 0.33  \\ 
1.10 2014 & 13& 0.045 & 21 &0.13  & 23& 0.43  \\ 
4.09.2015 & 14& 0.047 & 21 &0.12  & 24& 0.42  \\ 
26.10.2015& 15& 0.048 & 23 &0.13  & 24& 0.45  \\ 
12.02.2017& 13& 0.041 & 24 &0.14  & 24& 0.46  \\ 
13.06.2017& 15& 0.047 & 20 &0.14  & 23& 0.44  \\ 
3.08.2017 & 14& 0.041 & 19 &0.14  & 22& 0.41  \\ 
6.04.2018 & 14& 0.041 & 22 &0.14  & 23& 0.46  \\ 
\hline
\multicolumn{7}{l}{\footnotesize 1 -- the data from~[20], 2  -- the data from~[44].}
\end{tabular}   
\end{table}

It is also well known that the optical spectra of classical symbiotic and symbiotic nova stars as a rule
contain forbidden emissions. An example could be a cool peculiar supergiant PU\,Vul~[45, 46]. Symbiotic
stars are binary systems with a hot companion, UV radiation of which provides ionization of the gas envelope 
of the system and formation of the emission spectrum [47]. However, for the spectrum of a single
cold star V509\,Cas the presence of [NII] emissions are more difficult to explain. The likely mechanisms of excitation 
of these emissions in the spectrum of V509\,Cas are discussed in many publications since the discovery of these features. 
Sargent~[20] notes that the presence of forbidden [NII] emissions and complex emission-absorption H$\alpha$ and H$\beta$ 
profiles indicate the probable existence around V509\,Cas of a hot envelope. In the article~[48], devoted to the study 
of the chemical composition of the atmosphere of V509\,Cas, Luck adheres to the same position. However, Lambert et
al.~[5] do not find this version with a hot companion attractive, calling it only a passive spectator of the
cataclysms of the primary star. As long ago as in 1978 Lambert and Luck~[19] suggested several versions
on the mechanism of excitation of forbidden [NII] lines in the spectrum of V509\,Cas: dissipation of mechanical energy, 
ionization due to the UV radiation of hot stars of an HII-region in the volume of the
Cep\,OB1 association etc. Obviously, the principal in solving this problem is the spectrum of V509\,Cas in
the UV range. As the authors of~[30] showed, the UV spectrum of this star does not contain any features
and fully meets the expectations given the main parameters of this object.

A peculiar behavior of [NII] emissions is observed in the spectrum of the cold peculiar supergiant R\,CrB.
In a state of a deep brightness minimum, when the star is almost completely closed, optimal conditions
arise for the registration of a nebular spectrum of the envelope~[49]. An important moment is a long duration 
of the glare of the envelope in the [NII] emissions: as noted by the authors of~[50], at low density, the
nebula can remain ionized for up to 10$^4$ years.

In the spectrum of V509\,Cas, the half-width of the [NII] lines varies: $\delta$Vr\,=\,50\,km/s in September
1961~[20], 30\,km/s in 1967~[44]. The half widths for three lines [NII] 5755, 6548 and 6583\,\AA{} are measured
from our spectra, presented in Table\,2. It is necessary  to emphasize here that the half widths measured in
our spectra by five times exceed the half-width of the spectrograph’s point spread function. A reduction of the
emission half-widths in the spectra of V509\,Cas occurred after the 1996 observations and persisted over
the next 22 years of our observations. The intensities of the [NII] lines also vary with time: in 1967 the
equivalent widths W$_\lambda$ of the 6583 and 6548\,\AA{} lines were 0.13 and 0.045\,\AA{} respectively~[44], 
which is consistent with the theoretical intensity ratio 3:1 according to Osterbrock~[51]. According to our 
observations the behavior of line half-widths with time was accompanied by a synchronous change in their
intensities: in 1996 the equivalent widths increased by approximately 1.7 times, then the intensity stabilized.
Note that in the spectrum of the star the intensity measurements were carried out relative to the level
of the local continuous spectrum, so that at the constant emission rate in lines, their variability is also possible
due to the varying brightness of a star. According to the AAVSO database, V509\,Cas experiences a
slight brightness decrease in the past decade. Since the half-width also varies, it means that the velocity
dispersion changes in an optically thin envelope, i.e. the envelope is non-stationary.

\subsection{Radial velocity pattern}

We have fully identified the features in the spectrum of V509\,Cas. However, their huge number does not
allow us to give here a complete list of identified lines. Table~3 contains only those spectral features, the
positions of which are measured to study the velocity field. Expecting to find a probable velocity field 
stratification in the extended atmosphere of the hypergiant, we conducted a subsequent analysis of radial velocity
measurements, by combining the related spectral features into several groups. Group averages are presented
in Table\,1, where the dates of spectrum acquisition are indicated (in 1996 using the Lynx spectrograph,
further -- using the NES spectrograph), the working spectral intervals and heliocentric radial velocities Vr,
averaged for the spectral features of different nature and rounded to whole km/s.

The third column of Table\,1 lists the averaged velocities for the forbidden [FeII] emissions (10--13 features 
in the spectra with different wavelength ranges), while the fourth column gives them for the permitted
metal ion emissions. The subsequent columns for the features of other types:
\begin{itemize}
\item  such as the paired elevations above the continuous spectrum on the absorption wings (see the
FeII\,7462\,\AA{} profile in Fig.\,2) (over two hundred absorptions of SiII, ScII, TiII, CrII, FeII, YII, BaII). 
The listed values are given for the emissions in general, the central parts of which are superimposed by absorptions. 
The measurements are made from the lower parts of their profiles;
\item for the upper parts of the profiles  (wings) (fifth column) and FeI weak absorption cores (10--12 lines), 
   CaI (four lines) -- the sixth column;
\item for the cores of the components of the three strong FeII absorptions of the 42-nd multiplet in column~(7) 
  (as an example, Fig.\,2 shows the line profile of this multiplet FeII\,5169\,\AA{});

\item for the absorption components (in column~(8)) of the H$\alpha$ profile presented in Fig.\,3;

\item for three forbidden [NII] emissions 5755, 6548 and 6584\,\AA{} in column~(9);

\item for the four components of the NaI\,(1) profiles in columns (10)--(13). Figure~5 presents the 
  NaI\,5896\,\AA{} profile, averaged over the 2017 spectra; 
  
\item for a sample of diffuse interstellar DIB bands with the wavelengths $\lambda$: 5705, 5780, 5797, 5849, 6196,  
  the last 6203, 6284, 6379, 6613, 6660, 6672\,\AA{} --  in the last  column.
\end{itemize}

Owing to the high quality of the spectra and a long-term monitoring, we have obtained reliable conclusions in our 
study regarding the temporal behavior of the velocity field in the atmosphere of\,V509 Cas, which follow from the data 
in Table\,1. The proximity of the values presented in the first three columns, and their constancy in time indicate that 
the systemic velocity of V509\,Cas is close to Vsys=$-63$\,km/s. Exactly this value is described by a vertical broken 
line in Figs.\,2 and 3. Note that  the accepted value of the systemic velocity Vsys=$-63$\,km/s is not so far from the average 
velocity for the Cep\,OB1 association Vr=$-58.2$\,km/s~[52], a member of which is also
the hypergiant V509\,Cas.  Variations of velocity measured from the forbidden lines (3-d column of Table\,1) 
are minimal over time: Vr=$-$(62$\div$63)\,km/s. The velocity variability for the cores of FeII\,(42) 
absorptions is also minimal: Vr=$-$(84$\div$87)\,km/s, what testifies to the stability of the upper atmosphere. 
Note a similar behavior in time of the positions of forbidden emissions and FeII\,(42) core absorptions in 
the spectrum of the above-mentioned supergiant 3\,Pup~[32].

As follows from Table\,1 (column 6), velocity variability is the most prominent by the positions of strong absorption 
cores of the iron group ions: their  range from these absorptions is Vr=$-$(52$\div$71)\,km/s. This variability, 
which is well-illustrated by a panel with an NI\,7468 profile in Fig.\,2, in the case of a single star can be a 
manifestation of pulsations in the deep atmospheric layers, where the formation of this type of lines takes place. 
A pulsation variability of radial velocity in highly excited lines of neutral nitrogen NI with a full amplitude
of about 19\,km/s coinciding with our result was previously found by Sheffer and Lambert~[29]. They  found 
probable variability periods of 421 and 315 days, noting a constant, incessant profile instability. 
Recall that this type of pulsation variability with an amplitude of about 10\,km/s measured from the
weak and moderate-intensity symmetric absorptions is also inherent in the hypergiant $\rho$\,Cas~[9].

As follows from the data of Table\,1, forbidden [NII] emissions in the spectrum of V509\,Cas are
systematically shifted relative to the [FeII] emissions. However, this shift is stable, since for all times of our 
observations, the position of the [NII] emission remains in a small interval of velocities:
Vr(NIIf)=$-$(68$\div$69)\,km/s. A temporal constancy of this velocity is noted by Lambert et al.~[5],
taking the mean value of Vr(NIIf)=$-72$\,km/s for systemic velocity. Additional justification for this
choice of systemic velocity follows from the agreement of Vr(NIIf) with a velocity measured from
the emissions of a CO molecule, forming at the most outer layers of the extended atmosphere of
the hypergiant: Vr(CO)=$-$(72$\div$76)\,km/s~[5].  Later, the conclusion about the constancy of the
radial velocity measured from the forbidden emissions of [NII] in the optical spectra obtained during the long-term 
monitoring was also done by Sheffer and Lambert: Vr(NIIf)=$-72$\,km/s~[29] and Vr(NIIf)=$-69$\,km/s~[53].

A remnant of ejection of stellar matter located at a distance from the star can be considered a possible
source of radiation of forbidden [NII] emissions. Earlier, de~Jager~[54] proposed such  an explanation for
the occurrence of forbidden [NII] emissions in spectrum of V509\,Cas, indicating the formation of these
emissions in the envelope, the kinetic temperature in which is higher than the temperature of the star.
One has to keep in mind that the atmosphere of the star, and hence the matter ejected in the circumstellar 
neighborhood are largely enriched with nitrogen, because in the atmosphere of V509\,Cas the nitrogen
excess is over +3.5\,dex~[48]. Apparently, the so much altered chemical composition of the atmosphere 
explains the lack of expected emissions of oxygen ions, the relative abundance of which in the atmosphere of
the star is at least two orders of magnitude lower than that of nitrogen~[48].

\begin{figure}[t!]
\includegraphics[angle=0,width=0.35\textwidth,bb=40 110 335 595,clip]{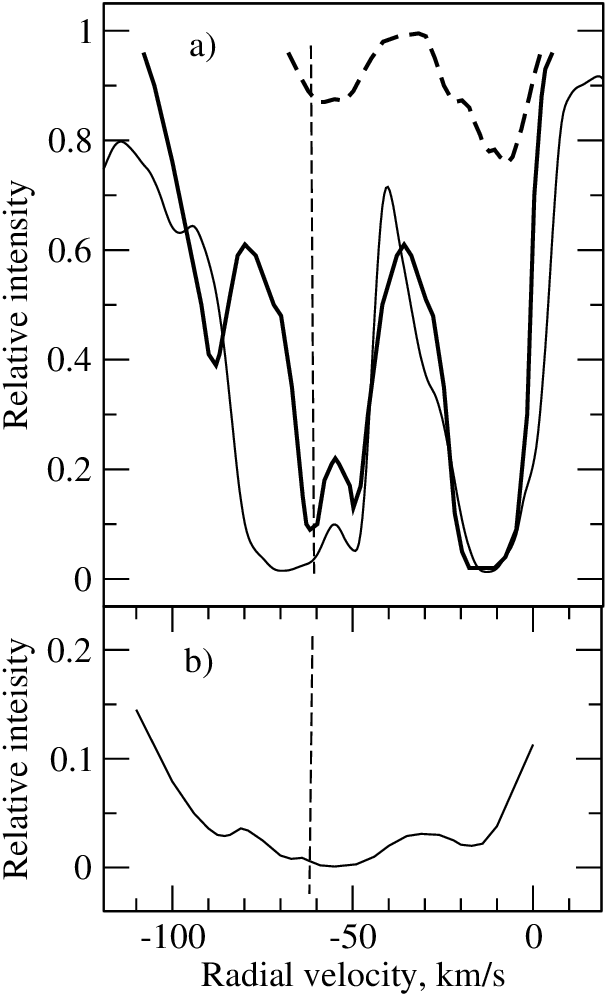}   
\caption{The profiles of selected features: (a) NaI~5896  D-lines in the spectrum of V509\,Cas are represented by
        the solid line, and in the spectrum $\rho$\,Cas -- by the thin line. The dashed line marks the profile of 
        KI~7699\,\AA{}  in the spectrum of V509\,Cas; (b) the CaII~3968\,\AA{}  line in the spectrum of V509\,Cas. 
        The vertical dashed line describes the accepted value of systemic velocity Vsys\,=\,$-63$\,km/s for V509\,Cas.} 
\end{figure}

\subsection{Multicomponent profile of NaI D-lines}

Multicomponent profiles of NaI D-lines (see Table\,1 and Fig.\,5a), in addition to the component near
the systemic velocity, also contain a narrow absorption with Vr$\approx -50$\,km/s and the poorly resolved
in our spectra saturated components in the velocity range of Vr$\approx -20; -30$\,km/s. A narrow absorption 
of Vr$\approx -50$\,km/s is formed in the interstellar medium and corresponds to the position of V509\,Cas
farther away than the Perseus Arm~[55, 56]. The positions of the circumstellar components in Fig.\,5, where
the NaI ~D-line profiles in the spectra of V509\,Cas and $\rho$\,Cas are compared~[9], having close galactic 
coordinates, are in a good agreement. An interstellar feature of Vr=$-13$\,km/s is registered in the spectrum of
the variable V354\,Lac~[57], the galactic coordinates of which are close to those for V509\,Cas and $\rho$\,Cas.

The structure of interstellar lines in the spectrum of V509\,Cas agrees well with its great remoteness, as well 
as with its extremely high luminosity, Mv$\approx -9^m$. We obtained the luminosity estimate
based on the equivalent width W$_\lambda$(OI\,7774)=2.35\,\AA{}  of the triplet OI~7774\,\AA{} 
in the spectrum of the star using the calibrations from~[14, 58]. For comparison note that in the spectrum 
of the hypergiant V1302\,Aql with its extreme luminosity near the Humphreys--Davidson limit~[59, 60], 
the equivalent  width of the oxygen triplet has a maximum value of W$_\lambda$(OI\,7774)=\,2.8\,\AA{}~[14].

The position of the NaI component with the velocity Vr=$-62; -63$\,km/s in Fig.\,5 is close
to the emissions and absorption wings, which indicates the formation of this component in the stellar
atmosphere. The position of the most shortwave depression in the NaI~D line profiles coincides with
the shortwave component of the FeII~(42) profiles: $-89$\,km/s in Fig.\,2. This agreement indicates
the formation of the most shortwave component in the outermost layers of the atmosphere close
to the circumstellar envelope. A broad long-wave component of the NaI~D-lines in the velocity range
of Vr$\approx -20; -30$\,km/s includes several unresolvable (at R=60\,000) interstellar features. The position
of one of them coincides with the reliably measured in the spectrum of V509\,Cas position of diffuse
interstellar bands (DIBs): of about $-14$\,km/s. 

The KI~7699\,\AA{} profile features, also presented in Fig.\,5, mostly correspond to those of the NaI~D-lines. 
However, interstellar bands of Vr$\approx -14$ and $-23$\,km/s, formed in the Local Arm are better identified 
in this profile. All these features of the interstellar and circumstellar origin are discernible in the core of 
the Ca\,II~3968\,\AA{} line in Fig.\,5b.

\section{Conclusions}  

Based on the data of spectral monitoring of the yellow hypergiant V509\,Cas, performed over 1996--2018 at 
the 6-m telescope with spectral resolution  R$\ge$60\,000, its kinematic state at various levels of extended 
atmosphere was studied in detail.
The proximity of velocities based on the permitted and forbidden emissions of metal ions, as well as their 
strict constancy in time led to the choice of systemic velocity of the hypergiant: Vsys=$-63$\,km/s.

No signs of binarity of the star were detected. 

For all our observations sets, the position of forbidden [NII] emissions, forming in the circumstellar
medium, corresponds  to the velocity Vr(NIIf)=$-$(68$\div$69)\,km/s. Therefore, [NII] emissions are systematically 
shifted by $-6$\,km/s relative to the metal ion emissions.  We made a conclusion that  after the 1996  
the variation of half widths and [NII] emission intensities (the lines have become narrower and more intense) 
and over the next 22 years of observations, these parameters did not vary. 

The velocities determined using the cores of wind absorptions FeII\,(42) are constant within the interval 
of Vr=$-$(84$\div$87)\,km/s, which indicates the stability of the uppermost layers of the atmosphere.

In general, we concluded on the stability of the hypergiant’s atmosphere excluding the layers closest 
to the photosphere. The velocity variability in range of  Vr=$-$(52$\div$71)\,km/s, determined by the positions 
of the cores of strong absorptions of the metal group ions may be a manifestation of pulsations in the deep  
atmospheric layers, where this type of lines is formed. 

\acknowledgements VGK thanks the Russian Science Foundation fort the partial financial support (project 14-50-00043).
ELCh and VEP thank the Russian Foundation for Basic Research for the partial support (projects nos.\,16-02-00587a and 
18-02-00029a). This work made use of the SIMBAD, SAO/NASA ADS,  AAVSO and VALD astronomical databases.

\newpage 

\begin{longtable}{ l  c|  c  c }
\caption{ Residual intensities r and heliocentric radial velocities Vr (km/s) for individual lines in the high resolution
                        spectrum of V509\,Cas}
\label{Lines}
\\ \hline
\endfirsthead
\hline
Ident & $\lambda$ & \hspace{0.5cm} r \hspace{0.3cm} &   Vr   \\
\hline
\endhead
\hline
\endfoot
\endlastfoot
Ident & $\lambda$ & r & Vr \\
\cline{1-4} 
YII\,(6)   &  3950.35  &   0.74 &  $-62 $  \\
VII\,(10 ) &  3951.96  &   0.66 &  $-58 $  \\
CaII\,(1)  &  3968.47  &   0.03 &  $-86 $  \\
           &           &   0.01 &  $-55 $  \\
           &           &   0.02 &  $-17 $  \\
H$\epsilon$&   3970.07 &   0.11 &  $-59 $  \\
TiII\,(11) &   3981.99 &   0.61 &  $-59 $  \\
TiII\,(11) &   3987.61 &   0.78 &  $-59 $  \\
FeII\,(126)&   4012.46 &   0.33 &  $-65 $  \\
TiII\,(11) &   4025.13 &   0.46 &  $-62 $  \\
TiII\,(87) &   4028.34 &   0.43 &  $-61 $  \\
FeI\,(43)  &   4045.81 &   0.50 &  $-60 $  \\
CrII\,(19) &   4051.97 &   0.75 &  $-60 $  \\
TiII\,(87) &   5053.83 &   0.44 &  $-58 $  \\
FeI\,(43)  &   4063.59 &   0.59 &  $-59 $  \\
NiII\,(11) &   4067.03 &   0.70 &  $-60 $  \\
FeI\,(43)  &   4071.74 &   0.66 &  $-61 $  \\
SrII\,(1)  &   4077.72 &   0.33 &  $-61 $  \\
H$\delta$  &   4101.74 &   0.13 &  $-58 $ \\
CrII\,(18) &   4110.99 &   0.75 &  $-59 $ \\
FeII\,(22) &   4124.78 &   0.81 &  $-57 $ \\
ZrII\,(41) &   4149.20 &   0.69 &  $-61 $ \\
TiII\,(21) &   4161.52 &   0.60 &  $-62 $ \\
TiII\,(105)&   4163.64 &   0.38 &  $-59 $ \\
TiII\,(105)&   4171.90 &   0.44 &  $-59 $ \\
FeII\,(27) &   4173.46 &   0.31 &  $-58 $ \\
FeII\,(28  &   4178.85 &   0.31 &  $-59 $ \\
SrII\,(1)  &   4215.52 &   0.39 &  $-60 $ \\
FeII\,(27) &   4233.17 &   0.20 &  $-57 $ \\
CrII,\,(31)&   4242.37 &   0.58 &  $-58 $ \\
ScII\,(7)  &   4246.83 &   0.22 &  $-59 $ \\
FeII\,(28) &   4258.15 &   0.62 &  $-61 $ \\
CrII\,(31) &   4261.92 &   0.64 &  $-59 $ \\
FeII\,(32) &   4278.15 &   0.78 &  $-59 $ \\
TiII\,(20) &   4287.88 &   0.56 &  $-60 $ \\
TiII\,(41) &   4290.21 &   0.27 &  $-60 $ \\
TiII\,(20) &   4294.10 &   0.25 &  $-61 $ \\
FeII\,(28) &   4296.57 &   0.50 &  $-60 $ \\
TiII\,(41) &   4300.04 &   0.19 &  $-58 $ \\
TiII\,(41) &   4307.89 &   0.30 &  $-61 $ \\
YII\,(5)   &   4309.63 &   0.82 &  $-63 $ \\
TiII\,(41) &   4312.86 &   0.30 &  $-58 $ \\
TiII\,(20) &   4337.92 &   0.25 &  $-59 $ \\
H$\gamma$  &   4340.47 &   0.14 &  $-55 $ \\
FeII\,(27) &   4351.77 &   0.29 &  $-58 $ \\
TiII\,(104)&   4367.66 &   0.63 &  $-59 $ \\ 
FeII\,(28) &   4369.40 &   0.74 &  $-59 $ \\
FeII\,(27) &   4385.38 &   0.42 &  $-59 $ \\
YII\,(5)   &   4398.02 &   0.82 &  $-56 $ \\
FeI\,(41)  &   4404.75 &   0.69 &  $-60 $ \\
TiII\,(51) &   4407.68 &   0.87 &  $-59 $ \\
ScII\,(14) &   4420.67 &   0.96 &  $-61 $ \\
TiII\,(19) &   4443.80 &   0.24 &  $-59 $ \\
TiII\,(19) &   4450.48 &   0.41 &  $-60 $ \\
FeII\,(26) &   4461.43  &  0.81 &  $-61 $ \\
TiII\,(40) &   4464.45  &  0.56 &  $-60 $ \\
TiII\,(31) &   4468.49  &  0.23 &  $-58 $ \\
TiII\,(40) &   4470.85  &  0.68 &  $-58 $ \\
FeII\,(37) &   4472.92  &  0.74 &  $-58 $ \\
MgII\,(4)  &   4481.22  &  0.44 &  $-60 $ \\
FeII\,(37) &   4491.40  &  0.49 &  $-60 $ \\
TiII\,(31) &   4501.27  &  0.25 &  $-59 $ \\
FeII\,(38) &   4508.28  &  0.37 &  $-59 $ \\
FeII\,(37) &   4515.34  &  0.41 &  $-59 $ \\
TiII\,(18) &   4518.33  &  0.82 &  $-58 $ \\
FeII\,(37) &   4520.22  &  0.41 &  $-58 $ \\
FeII\,(38) &   4522.63  &  0.32 &  $-58 $ \\
TiII\,(82) &   4529.49  &  0.67 &  $-61 $ \\
CrII\,(39) &   4539.62  &  0.91 &  $-58 $ \\
FeII\,(38) &   4541.52  &  0.56 &  $-60 $ \\
TiII\,(60) &   4544.02  &  0.86 &  $-61 $ \\
TiII\,(30) &   4545.14  &  0.83 &  $-62 $ \\
CrII\,(44) &   4558.64  &  0.41 &  $-59 $ \\
TiII\,(50) &   4563.76  &  0.26 &  $-58 $ \\
CrII\,(39) &   4565.77  &  0.84 &  $-61 $ \\
TiII\,(82) &   4571.97  &  0.21 &  $-59 $ \\
FeII\,(38) &   4576.34  &  0.56 &  $-60 $ \\
FeII\,(38) &   4583.83  &  0.27 &  $-58 $ \\
CrII\,(44) &   4588.20  &  0.47 &  $-60 $ \\
CrII\,(44) &   4592.05  &  0.69 &  $-59 $ \\
VII\,(56)  &   4600.19  &  0.93 &  $-59 $ \\
CrII\,(44) &   4616.62  &  0.72 &  $-59 $ \\
CrII\,(44) &   4618.82  &  0.55 &  $-61 $ \\
FeII\,(38) &   4620.51  &  0.68 &  $-60 $ \\
FeII\,(37) &   4629.33  &  0.38 &  $-60 $ \\
CrII\,(44) &   4634.07  &  0.61 &  $-59 $ \\
FeII\,(37) &   4666.75  &  0.67 &  $-59 $ \\
TiII\,(49) &   4708.67  &  0.80 &  $-61 $ \\
FeII\,(43) &   4731.47  &  0.61 &  $-61 $ \\
$[FeII]$20F&   4774.72  &  1.03 &  $-64 $ \\
TiII\,(92) &   4779.98  &  0.66 &  $-60 $ \\
TiII\,(17) &   4798.53  &  0.86 &  $-59 $ \\
TiII\,(92) &   4805.09  &  0.54 &  $-60 $ \\
CrII\,(30) &   4812.35  &  0.84 &  $-60 $ \\
$[FeII]$20F&   4814.53  &  1.05 &  $-62 $ \\
CrII\,(30) &   4824.14  &  0.53 &  $-60 $ \\
CrII\,(30) &   4836.24  &  0.83 &  $-60 $ \\
CrII\,(30) &   4848.25  &  0.61 &  $-61 $ \\
H$\beta$   &   4861.33  &  0.21 &  $-67 $ \\
TiII\,(114)&   4874.01  &  0.78 &  $-62 $ \\
CrII\,(30) &   4876.40  &  0.65 &  $-58 $ \\
$[FeII]$4F &   4889.62  &  1.09 &  $-58 $ \\
FeII\,(36) &   4893.81  &  0.93 &  $-59 $ \\
$[FeII]$20F&   4905.34  &  1.03 &  $-61 $ \\
TiII\,(114)&   4911.19  &  0.71 &  $-60 $ \\
FeII\,(42) &   4923.92  &  0.21 &  $-56 $ \\
BaII\,(1)  &   4934.08  &  0.80 &  $-60 $ \\
$[FeII]$20F&   4947.37  &  1.02 &  $-63 $ \\
$[FeII]$20F&   4950.74  &  1.02 &  $-61$  \\
FeII\,(36) &   4993.35  &  0.81 &  $-60$  \\
TiII\,(71) &   5013.69  &  0.84 &  $-62$  \\
FeII\,(42) &   5018.44  &  0.49 &  $-88$  \\
           &            &  0.22 &  $-55$  \\
SiII\,(5)  &   5041.03  &  0.86 &  $-62$  \\
SiII\,(5)  &   5056.06  &  0.83 &  $-65$  \\
TiII\,(113)&   5072.30  &  0.85 &  $-60$  \\
YII\,(20)  &   5087.42  &  0.88 &  $-63$  \\
ZrII\,(95) &   5112.27  &  0.94 &  $-59$  \\
FeII\,(35) &   5120.34  &  0.94 &  $-59$  \\
FeII       &   5123.19  &  0.95 &  $-61$  \\
TiII\,(86) &   5129.16  &  0.67 &  $-61$  \\
FeII\,(35) &   5132.67  &  0.92 &  $-60$  \\
FeII\,(35) &   5146.12  &  0.89 &  $-62$  \\
TiII\,(70) &   5154.08  &  0.68 &  $-61$  \\
$[FeII]$   &   5158.78  &  1.06 &  $-62$  \\
FeII\,(42) &   5169.03  &  0.45 &  $-88$  \\
           &            &  0.35 &  $-70$  \\
           &            &  0.21 &  $-54$  \\
MgI\,(2)   &   5172.69  &  0.68 &  $-59$  \\
MgI\,(2)   &   5183.61  &  0.63 &  $-60$  \\
TiII\,(86) &   5185.91  &  0.70 &  $-61$  \\
TiII\,(70) &   5188.69  &  0.49 &  $-61$  \\  
FeII\,(49) &   5197.58  &  0.48 &  $-60$  \\
YII\,(20)  &   5200.41  &  0.92 &  $-63$  \\
YII\,(20)  &   5205.73  &  0.86 &  $-61$  \\
TiII\,(70) &   5226.55  &  0.54 &  $-60$  \\
FeII\,(49) &   5234.62  &  0.45 &  $-60$  \\
CrII\,(43) &   5237.32  &  0.67 &  $-60$  \\
ScII\,(26) &   5239.82  &  0.84 &  $-61$  \\
FeII       &   5254.93  &  0.79 &  $-60$  \\
FeII\,(49) &   5256.93  &  0.91 &  $-59$  \\
FeII\,(48) &   5264.80  &  0.74 &  $-59$  \\
$[FeII]$18F&   5273.35  &  1.10 &  $-62$  \\
FeII\,(49) &   5276.00  &  0.42 &  $-58$  \\
FeII\,(41) &   5284.10  &  0.66 &  $-60$  \\
CrII\,(24) &   5305.86  &  0.88 &  $-59$  \\
CrII\,(43) &   5308.42  &  0.90 &  $-60$  \\
CrII\,(43) &   5313.58  &  0.82 &  $-59$  \\
FeII\,(48) &   5316.66  &  0.36 &  $-59$  \\
FeII\,(49) &   5325.56  &  0.77 &  $-59$  \\
FeI\,(15)  &   5328.04  &  0.88 &  $-56$  \\
$[FeII]$19F&   5333.65  &  1.04 &  $-61$  \\
CrII\,(43) &   5334.87  &  0.83 &  $-58$  \\
ZrII\,(115)&   5350.09  &  0.97 &  $-61$  \\
FeII\,(48) &   5362.87  &  0.57 &  $-61$  \\
FeI\,(15)  &   5371.50  &  0.95 &  $-61$  \\
$[FeII]$19F&   5376.45  &  1.04 &  $-61$  \\
TiII \,(69)&   5381.03  &  0.80 &  $-60$  \\
CrII\,(23) &   5407.62  &  0.95 &  $-60$  \\
TiII\,(69) &   5418.78  &  0.85 &  $-61$  \\
CrII\,(23) &   5420.93  &  0.93 &  $-64$  \\
FeII       &   5425.25  &  0.80 &  $-61$  \\
FeI\,(15)  &   5429.70  &  0.96 &  $-60$  \\
FeI\,(15)  &   5455.61  &  0.96 &  $-58$  \\
CrII\,(50) &   5478.37  &  0.88 &  $-63$  \\
TiII\,(68) &   5490.69  &  0.96 &  $-62$  \\
FeII\,(55) &   5534.86  &  0.66 &  $-60$  \\
$[FeII]$39F&   5551.31  &  1.02 &  $-61$  \\
FeI\,(686) &   5615.64  &  0.94 &  $-64$  \\
FeII\,(57) &   5627.49  &  0.96 &  $-59$  \\
ScII\,(29) &   5640.98  &  0.90 &  $-61$  \\
YII\,(38)  &   5662.95  &  0.88 &  $-64$  \\
ScII\,(29) &   5667.15  &  0.93 &  $-62$  \\
ScII\,(29) &   5669.03  &  0.90 &  $-62$  \\
ScII\,(29) &   5684.19  &  0.89 &  $-63$  \\
NaI\,(6)   &   5688.21  &  0.95 &  $-60$  \\
DIB        &   5705.20  &  0.96 &  $-22$  \\
$[NII]$3F  &   5754.64  &  1.08 &  $-67$  \\
DIB        &   5780.37  &  0.84 &  $-20$  \\
FeII\,(164)&   5823.15  &  0.99 &  $-64$  \\
FeII\,(182)&   5835.49  &  0.99 &  $-62 $ \\
DIB        &   5849.80  &  0.97 &  $-16 $ \\
BaII\,(2)  &   5853.68  &  0.97 &  $-63 $ \\
CaI\,(47)  &   5857.46  &  0.99 &  $-62 $ \\
NaI\,(1)   &   5889.95  &  0.21 &  $-90 $ \\
           &            &  0.09 &  $-62 $ \\
           &            &  0.16 &  $-50 $ \\
           &            &  0.02 &  $-14 $ \\
NaI\,(1)   &   5895.92  &  0.36 &  $-90 $ \\
           &            &  0.11 &  $-62 $ \\
           &            &  0.20 &  $-51 $ \\
           &            &  0.03 &  $-14 $ \\
SiII\, (4) &   5978.93  &  0.93 &  $-65 $ \\
FeII\,(46) &   5991.37  &  0.91 &  $-60 $ \\
FeII\,(46) &   6084.10  &  0.94 &  $-61 $ \\
FeII\,(46) &   6113.32  &  0.96 &  $-63 $ \\
CaI\,(3)   &   6122.22  &  0.97 &  $-65 $ \\
BaII\,(2)  &   6141.72  &  0.87 &  $-63 $ \\
FeII\,(74) &   6147.74  &  0.80 &  $-62 $ \\
CaI\,(3)   &   6162.18  &  0.97 &  $-64 $ \\
DIB        &   6195.96  &  0.93 &  $-14 $ \\
FeII\,(162)&   6199.19  &  0.99 &  $-59 $ \\
DIB        &   6203.08  &  0.94 &  $-17 $ \\
FeII\,(74) &   6238.39  &  0.81 &  $-60 $ \\
ScII\,(28) &   6245.62  &  0.92 &  $-61 $ \\
FeII\,(74) &   6247.55  &  0.70 &  $-59 $ \\
DIB        &   6283.85  &  0.81 &  $-22 $ \\
SiII\,(2)  &   6347.10  &  0.60 &  $-59 $ \\
FeII\,(40) &   6369.47  &  0.93 &  $-60 $ \\
SiII\,(2)  &   6371.36  &  0.67 &  $-59 $ \\
DIB        &   6379.29  &  0.94 &  $-15 $ \\
FeI\,(168) &   6393.61  &  0.99 &  $-65 $ \\
FeI\,(816) &   6400.01  &  0.99 &  $-66 $ \\
FeI\,(816) &   6411.65  &  0.97 &  $-57 $ \\
FeII\,(74) &   6416.93  &  0.84 &  $-59 $ \\
CaI\,(18)  &   6439.08  &  0.98 &  $-65 $ \\
FeII       &   6442.95  &  0.97 &  $-65 $ \\
FeII\,(74) &   6456.38  &  0.60 &  $-61 $ \\
BaII\,(2)  &   6496.91  &  0.64 &  $-56 $ \\
$[NII]$1F  &   6548.03  &  1.12 &  $-68 $ \\
H$\alpha$  &   6562.81  &  0.52 &  $-96 $ \\
           &            &  1.27 &  $-22 $ \\
$[NII]$1F  &   6583.45  &  1.40 &  $-70 $ \\
DIB        &   6613.56  &  0.89 &  $-16 $ \\
DIB        &   6660.64  &  0.96 &  $-11 $ \\
DIB        &   6672.15  &  0.98 &  $-10 $ \\
TiII\,(112)&   6717.91  &  0.93 &  $-57 $ \\
SI\,(8)    &   6748.79  &  0.97 &  $-57 $ \\
SI\,(8)    &   6757.16  &  0.96 &  $-52 $ \\
$[FeII]$14F&   7155.14  &  1.08 &  $-64 $ \\
$[CaII]$1F &   7291.46  &  1.24 &  $    $ \\
$[FeII]$14F&   7388.16  &  1.03 &  $-63 $ \\
MnII\,(4)  &   7415.78  &  0.88 &  $-52 $ \\
FeII(73)   &   7462.39  &  0.80 &  $-54 $ \\
NI\,(3)    &   7468.31  &  0.70 &  $-51 $ \\
FeII\,(72) &   7479.69  &  0.94 &  $-53 $ \\
FeII\,(73) &   7515.79  &  0.91 &  $-54 $ \\
FeII\,(72) &   7533.36  &  0.91 &  $-54 $ \\
KI     (1) &   7698.97  &  0.86 &  $-61 $ \\
           &            &  0.86 &  $-56 $ \\ 
           &            &  0.75 &  $-11 $ \\
FeII\,(73) &   7711.71  &  0.76 &  $-55 $ \\
DIB        &   7721.85  &  0.96 &  $-11 $ \\
OI\,(1)    &   7771.94  &  0.40 &  $-55 $ \\
OI\,(1)    &   7774.17  &  0.41 &  $    $ \\
OI\,(1)    &   7775.39  &  0.48 &  $-54 $ \\
MgII\,(8)  &   7896.37  &  0.81 &  $-55 $ \\
H\,(P20)   &   8392.40  &  0.63 &  $-58 $ \\
H\,(P19)   &   8413.32  &  0.65 &  $-57 $ \\
H\,(P18)   &   8437.96  &  0.64 &  $-55 $ \\
OI\,(4)    &   8446.38  &  0.39 &  $-52 $ \\
H\,(P17)   &   8467.26  &  0.58 &  $-53 $ \\
\hline
\end{longtable}

\end{document}